\pdfoutput=1

\documentclass[a4paper,10pt,twoside]{article}
\usepackage[T1]{fontenc}

\usepackage{multirow}
\usepackage[utf8x]{inputenc}
\usepackage{graphicx}
\usepackage{url}
\usepackage{setspace}
\usepackage{subfig}
\usepackage{hyperref} % For hyperlinks in the PDF

\usepackage{abstract} % Allows abstract customization
\usepackage{titlesec} % Allows customization of titles
\renewcommand\thesection{\Roman{section}} % Roman numerals for the sections
\renewcommand\thesubsection{\roman{subsection}} % roman numerals for subsections
\titleformat{\section}[block]{\large\scshape\centering}{\thesection.}{1em}{} % Change the look of the section titles
\titleformat{\subsection}[block]{\large}{\thesubsection.}{1em}{} % Change the look of the section titles

\usepackage{titling} % Customizing the title section

\usepackage[hmarginratio=1:1,top=32mm,columnsep=20pt]{geometry} % Document margins
\newcommand{\comentario}[1]{}

\begin{document}
% \doublespacing
% \input{reviews.txt}

\title{Decanting the Contribution of Instruction Types and Loop Structures in the Reuse of Traces}

\author{%
\textsc{Andrey M. Coppieters, Sheila de Oliveira, Felipe M. G. França}\\ [1ex]
\normalsize Program of Systems and Computer Engineering\\
\normalsize Federal University of Rio de Janeiro\\
\normalsize PO Box 68511\\
\normalsize Rio de Janeiro, RJ/21941-972, Brazil\\
\normalsize \href{mailto:andrey.coppieters@tvglobo.com.br}{andrey.coppieters@tvglobo.com.br}, \href{mailto:sdomally@gmail.com}{sdomally@gmail.com}, \href{mailto:felipe@cos.ufrj.br}{felipe@cos.ufrj.br}
~\\
~\\
~\\
\textsc{Maurício L. Pilla}\\ [1ex]
\normalsize Graduate Program in Computer Sciences\\
\normalsize Federal University of Pelotas\\
\normalsize R. Gomes Carneiro, 1\\
\normalsize Pelotas, RS/96010-610, Brazil\\
\normalsize \href{mailto:pilla@inf.ufpel.edu.br}{pilla@inf.ufpel.edu.br}
~\\
~\\
~\\
\textsc{Amarildo T. da Costa}\\ [1ex]
\normalsize Military Institute of Engineering,\\
\normalsize Praça Gen. Tibúrcio, 80\\
\normalsize Rio de Janeiro, RJ/22290-270, Brazil\\
\normalsize \href{mailto:amarildo@cos.ufrj.br}{amarildo@cos.ufrj.br}
}

\date{}
\maketitle

\begin{abstract}
Reuse has been proposed as a microarchitecture-level mechanism to reduce the amount of executed instructions, collapsing dependencies and freeing resources for other instructions. Previous works have used reuse domains such as memory accesses, integer or not floating point, based on the reusability rate. However, these works have not studied the specific contribution of reusing different subsets of instructions for performance. In this work, we analysed the sensitivity of trace reuse to instruction subsets, comparing their efficiency to their complementary subsets. We also studied the amount of reuse that can be extracted from loops. Our experiments show that disabling trace reuse outside loops does not harm performance but reduces in 12\% the number of accesses to the reuse table. Our experiments with reuse subsets show that most of the speedup can be retained even when not reusing all types of instructions previously found in the reuse domain.
\end{abstract}
% }

\pagebreak

% \begin{keyword}
% {Value Reuse; Computer Architectures; Instruction Sets.}
% \end{keyword}

%-------------------------------------------------------------------------
\section{Introduction}

A strong trend in the design of modern processors regards energetic efficiency of chips. Gate switching was one of the main components of power budgets, but in the last years CMOS technologies have been reducing the size of gates to a point where even leakage current can play an important role. Therefore, even microarchitecure mechanisms for general-purpose processors must take into account optimizations to reduce the amount of spent energy.

Reuse has been proposed as a microarchitecture-level mechanism to reduce the amount of executed instructions, collapsing dependencies and freeing resources for other instructions, thus improving performance and energy consumption. Previous works have considered reuse domains such as memory accesses, integer, or any instruction other than floating point ones, based on the reusability rate. However, these studies have not studied the specific contribution of reusing different subsets of instruction types for performance.

In this work, we intend to {\bf improve the understanding of how the performance gains of trace reuse are distributed in the code execution}. This knowledge may be used to help reducing the number of accesses to the reuse table and reduce energy consumption, while keeping good performance.

This paper presents these main contributions:

\begin{enumerate}
\item The contribution of reuse inside and outside loops is presented, comparing performance and reuse table accesses for those to reusing in the entire code;
\item The sensitivity of trace reuse to subsets such as logical and arithmetic, branches, and memory address calculations is analysed, comparing their efficiency with their complementary subsets;
\item A comparison of trade-offs for memory budgets is discussed, where performance of reuse is compared to a similar area in first-level caches.
\end{enumerate}

Our experiments show that disabling trace reuse outside loops does not harm performance but reduces in 12\% the number of accesses to the reuse table. In the experiments with subsets, reusing only logical and arithmetic instructions achieves 92.8\% of the speedup of reusing the entire reuse domain, reusing only 42.8\% of the instructions. Reusing only branches produces 90.3\% of the speedup, reusing only 14.5\% of the instructions. Measurements of simulated energy spent in the reuse table support that reducing the number of accesses is much more important than reducing its size. Memory budget comparisons showed that doubling first-level caches is less effective than using the same area for reuse tables.

This paper is divided in following sections. First,  related work is discussed in Section~\ref{sec:previous}. Then, the Reuse through Speculation on Traces~(RST) architecture is revisited in Section~\ref{sec:rst}. Section~\ref{sec:method} briefly presents the simulation methodology and benchmarks used in this work. Section~\ref{sec:loops} presents our study about reuse inside and outside loops, while Section~\ref{sec:instructions} shows the contribution of different subsets of the reuse domain to performance and in-chip memory accesses. Finally, Section~\ref{sec:conclusion} presents our concluding remarks and discussion of future works.

\section{Related Work}
\label{sec:previous}

Many different mechanisms based on value reuse have been proposed. The reuse
granularity includes instructions~\cite{ROT00micro,SOD98micro}, expressions and
invariants~\cite{MOL99super}, basic blocks~\cite{HUA00comp},
traces~\cite{COS99tr,COS00pact,GON99icpp}, as well as instruction blocks and sub-blocks of
arbitrary size~\cite{HUA00pact}. The techniques vary in terms of
their dependence on hardware and compiler support~\cite{HUA00pact,WU01isca}. %HUA99hpca

In most studies, memory accesses are not reused because of side effects and aliasing disambiguation difficulties. One approach to implement load and store reuse is to manage registers as a
level in the memory hierarchy~\cite{OND01super}. Another approach employs instruction
reuse to exploit both same instruction and different instruction
redundancy~\cite{YAN00icpp}.

%Load value
%prediction~\cite{LIP96micro} is a specialization where only loads are predicted,
%motivated by their long latencies and high value locality. Another
%version~\cite{TUL99isca} uses the value stored in a register as the prediction,
%without requiring extra tables.

Many variations on value prediction have been proposed, including two-level value
prediction~\cite{WAN97micro}, hybrid value prediction~\cite{WAN97micro,SAT98micro2},
and others, many of which were inspired by branch prediction. Value prediction based on correlation~\cite{WAN97micro,SAZ97micro} uses global
information about the path in selecting  predictions. Prediction of
multiple values for traces~\cite{SAT98micro2} may be
done for only the live-out values in a trace, reducing necessary
bandwidth in the predictor. Speculation control~\cite{GRU98isca} is used to balance the
benefits of speculation against other possibilities. Confidence mechanisms~\cite{CAL99isca} are employed to improve value prediction by restricting prediction of unpredictable values. Past work that explored the limits of value prediction with trace reuse has shown there is much potential for performance improvement~\cite{PIL03sbac}.

%Another work~\cite{TUN01hpca} suggests
%finding and predicting chains of dependent instructions in the
%critical path to optimize performance.
Wu~et~al.~\cite{WU01isca} proposed speculative reuse but requires compiler support. RST has the advantage of being independent of special compiler support, ISA changes, and extra execution engine for misprediction recovery.

Huang, Choi and Lilja~\cite{HUA99tr} proposed a scheme that uses a Speculative Reuse
Cache (SRC) and Combined Dynamic Prediction (CDP) to exploit value reuse and prediction.  During execution, a chooser picks a prediction from the value predictor or a speculatively reusable value from SRC.  With this approach, they have achieved a speedup of 10\% in a 16-wide, 6-stage superscalar architecture, with 128 KB of storage for the CDP. RST has higher overall speedups and requires less storage~\cite{PIL03sbac,PIL06sbac,PIL07ijhpca}.

Liao and Shieh~\cite{LIA02apccsa} combined instruction reuse and value
prediction, using a reuse buffer and a value prediction
table, that are accessed in parallel. If inputs are unavailable, the value predictor is used. For a 6-stage pipeline, they achieved an average speedup of 9\%. Unlike RST, this approach speculatively reuses instructions, rather than traces, and it requires two tables.

The Contrail architecture~\cite{CONTRAIL2003} used two threads, one with speculative execution running on low-latency functional unities, and another thread validating results predicted by the first thread. The speculation thread aims to transform critical paths into non-critical paths, and the execution on slower functional unities allow for reductions of up to 40\% on energy dissipation. RST does not require an extra thread to validate results from speculation, making it simpler to implement. Trace speculation recovery uses a mechanism similar to those used for branch misprediction.

Our previous work in~\cite{PIL06sbac} studied the upper bound limits of RST in a superscalar architecture with oracle prediction. In~\cite{PIL07ijhpca}, we did an extensive study on upper bound limits for feasible architectures, proposing a version with a unified reuse table to further reduce die area requirements. In this work, we intended to verify the impact of reducing the reuse domain or by reusing only instructions inside loops in overall performance.

Pratas et al~\cite{Pratas:2008:ccf} developed a new constrained loop unrolling technique for reusing instruction inside loops in low power architectures. A controller was designed to identify reusable instructions within loops. This technique is more closely related to Trace Caches~\cite{Rotenberg:1996:micro} than trace reuse, as both do not store operands but just the instruction themselves. However, \cite{Pratas:2008:ccf} does not require an extra buffer, using the reorder buffer as source of instructions.

Tsumura et al~\cite{Tsumura:2007:pdcn} identifies reusable regions by tracking subroutine call and return instructions, or by backward branches in the case of loops. This architecture uses threads running on other cores to validate speculative reuse. In our paper, we also tracked loops by observing backward branches, although subroutine calls were not analysed. Oda et al~\cite{Oda:2011:icnc} extended Tsumura's previous work by improving the efficiency of input memoization.

Tsai and Chen~\cite{Tsai:2011:vlsi} proposed a Trace Reuse~(TR) Cache to reduce energy consumption in embedded systems. Instead of reusing the computations, TR only reuses the decoded sequences of retired instructions, thus avoiding cache misses and improving overall efficiency. However, this mechanism was evaluated for simple embedded processors and may probably not scale well for use in out-of-order, more complex processors.

Rahimi et al.~\cite{Rahimi:2013:tcs} used spatial memoization to reduce the costs of recovering errors in a SIMD architecture. An interesting feature of this work is the absence of reuse tables, where reuse occurs from an instruction running on a strong lane to the ones on the remaining weak lanes. Experimental results showed that 62\% of recovery of errant instructions is avoided using their technique. The authors extended their work for floating points in GPUs~\cite{Rahimi:2014:date}, with results of energy saving from 8~to~28\% for error correction.

\comentario{Is there any other paper that we should cite here to address Reviewer \#1's requests?}

\section{The RST Architecture}
\label{sec:rst}

RST~(Reuse through Speculation on Traces) improves trace reuse and hides true data dependencies by allowing traces to be {\em (i)}~{\bf regularly reused} when all inputs are ready and match previously stored input values  or {\em (ii)}~{\bf speculatively reused} when there are unknown trace inputs. Therefore, traces that could not be reused in previous approaches may be reused to exploit their redundancy.

Value reuse is non-speculative. After the input values of an instruction are verified against previous values and a match is found, the instruction can be reused without further execution. Resources are never wasted due to reuse and are available to other instructions. In trace reuse, a sequence of instructions--possibly spanning multiple branches--is the granularity of reuse.  Input values to a trace are compared against previous values to check for a reuse opportunity, and when the inputs match the previous values, previously recorded outputs are used to update the architecture state. When a trace is reused, its outputs are written to registers, instructions are skipped, and true dependencies are collapsed.

Traces are created during runtime  based on sequences of instructions in a {\em reuse domain}, which is the set of instruction  classes that are allowed to be reused (e.g., integer operations,  branches, and address calculations for loads and stores). Each trace has an input and an output context to store live-in and live-out values. The reuse domain is constituted by integer instructions without side effects.  Floating-point instructions are not reused because they show little redundancy~\cite{COS00pact} and because they would require larger registers to be stored in the reuse table. The length of traces may be limited by the number of live-in and live-out values that can be stored, by the number of branches, and by the occurrence of instructions outside of the reuse domain.  In loads and stores, the address calculation is split from the actual memory access, and the address calculation can be reused. System calls are  not reused because they require a mode change in the processor status.

\begin{figure}[!ht]
\begin{center}
\includegraphics[width=0.7\textwidth]{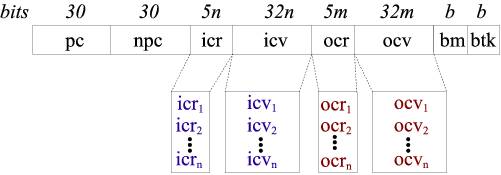}
\caption{Structure of an entry in {\em Memo\_Table\_T}.}
\label{fig:trace_memoization}
\end{center}
\end{figure}

Figure~\ref{fig:trace_memoization} shows the structure of an entry in the reuse table ({\em Memo\_Table\_T})~\cite{PIL07ijhpca}. The address of the first instruction of a trace is stored in {\em pc}, and the address of the next instruction after the trace is stored in {\em npc}. The {\em npc} field is used to set the PC and to skip the instructions belonging to the trace when it is reused. Fields {\em icv} and {\em icr} store the  input values and register indexes, while {\em ocv} and {\em ocr} store the output values and register indexes. {\em bm} is a bitmap to  mark branches that occur inside the trace, while {\em btk} bits are set when branches are taken. These two last fields are used to update the branch prediction mechanism.

In this work, the unified reuse table proposed in~\cite{PIL07ijhpca} is used for the first set of experiments. This implementation uses a single table to store both isolated instructions and traces, reducing size requirements and also the number of table accesses, with performance results similar to using two tables. Besides, a unified table allows for speculative reuse of single instructions.

The second set of experiments rely on two tables, one for instruction reuse ({\em Memo\_Table\_G}), and {\em Memo\_Table\_T} for traces with two or more instructions. Details on trace construction and reuse can be found in~\cite{PIL03sbac,PIL06sbac,PIL07ijhpca}.
%
% More details about how RST works can be found in~\cite{PIL06sbac,PIL07ijhpca}.

\section{Simulation Methodology}
\label{sec:method}

For the studies presented in this paper, two different set of simulation tools based on modifications of {\em sim-rst}~\cite{PIL07ijhpca} were used. {\em sim-rst} is an out-of-order simulator based on {\em sim-outorder} from SimpleScalar~\cite{BUR97tr}, implementing in details a speculative trace reuse technique.

\comentario{
  Addressing Reviewer \#2's comments on methodology and simulation: We are afraid that it is not possible to reimplement the simulator in another platform in a timely way. These results are part of the work of two masters students that concluded their studies before the paper submission, and the implementation of the required modifications on the simulator was neither trivial nor straightforward.}

\comentario{ Simulating reuse on these simulators requires extensive modification and is heavily attached to both the ISA and the implementation of the simulator.
}

\comentario{
A quick search in Google Scholar shows 252 papers that cite/use SimpleScalar since last year, hence we assert that our results using this simulator are still meaningful. The same goes for the use of the SPEC CPU 2000, with 190 hits.
}

\comentario{Although it is 14 years old (it was less than 12 years old when we finished writing this paper), newer versions of SPEC CPU require the execution of a much larger number of dynamic instructions and thus is not appropriate for detailed architecture simulation such as we did.
}

The first set included {\em sim-rst-loop} and {\em sim-rst-outofloop}. They both use backward branches to dynamically detect loop structures. The former simulator enables reuse only when such structures are detected, and disables it when the end of a loop is detected. The later does the reverse, enabling reuse only outside loops. These simulators were used in the first set of experiments to determine how much the natural redundancy in loops contributed to performance.

The second set of simulators based on {\em sim-rst} enabled or disabled the reuse of certain subsets of the reuse domain. These simulators were used to investigate the sensibility of trace reuse to instruction type.

The benchmarks {\tt bzip2}, {\tt gcc}, {\tt gzip}, {\tt mcf}, {\tt parser}, and {\tt vortex} from \mbox{{SPEC}~{CPU}~2000}~benchmarks were simulated in all experiments. These benchmarks were were found to be an interesting subset of SPEC CPU 2000 and simulating the entire set would be costly and probably would not bring further insights on the aspects that we were studying, as each simulation with a given configuration takes more than a day. Some of those benchmarks are even common to the SPEC CPU 2006 set.

We did not use sampling, because might break dynamic instruction flows that would produce redundancy, which was the aspect that we were mostly interested. Our previous study on limits of trace reuse used a similar methodology with very good results. In the first set of experiments,  each benchmark is run for 1.5~billion instructions, from which 500~million instructions were fast-forwarded and the remaining 1~billion instructions were simulated. In the second set of experiments, from 1.7~to~2.5~billion instructions were simulated for each benchmark.

The baseline architecture configuration is summarized in Table~\ref{tab:configs}.

\begin{table}[!ht]
%\vspace{-0.5cm}
\caption{Architecture configuration}
\label{tab:configs}
%\vspace{-1mm}
\begin{small}
\begin{center}
\begin{tabular}{|l|c|}
\hline
{\bf Parameter} & {\bf Value}\\
\hline\hline
Pipeline width & 4 (2 ALUs, 2 memory, 1 mult)\\
\hline
\multirow{2}{*}{Pipeline depth} & 19 (4 fetch, 4 decode, 2 dispatch,\\
& 5 issue, 1 execute, 2 writeback, 1 commit)\\
\hline
IFQ, RUU, LSQ & 16 instructions, 128 entries, 64 entries\\
\hline
Branch predictor & 2-level\\
\hline
First level & 13-bit register (xored with PC)\\
\hline
Second level & 8192 entries\\
\hline
BTB & 4096~entries, 2-way\\
\hline
Instruction set & PISA\\
\hline
L1 cache & 32KB I \& D, 4-assoc, 64B lines, 1 cycle hit\\
\hline
L2 cache & 512KB, 8-assoc, 256B lines, 5 cycles hit\\
\hline
L3 cache & 2MB, 8-assoc, 256B lines, 20 cycles hit\\
\hline
Memory & 200 cycles first chunk, 20~cycles next chunk\\
\hline
\end{tabular}
%\vspace{-0.5cm}
\end{center}
\end{small}
\end{table}

In the first set of experiments, the same configuration with a single reuse table entry used in~\cite{PIL07ijhpca} was employed, because of the good balance between performance and table size. For the second set of experiments, larger input and output scopes were employed to avoid limiting reuse of longer traces. Two separated tables for instruction and trace reuse, {\em Memo\_Table\_G} and {\em Memo\_Table\_T}, were used. Table~\ref{tab:reuse} sumarizes the configurations for each experiment set.

\begin{table}[!ht]
 \caption{Reuse table configurations.}
\label{tab:reuse}
\begin{center}
 \begin{small}
\begin{tabular}{|l|c|c|}
 \hline
{\bf Parameter}  & {\bf First set} & {\bf Second set}\\
\hline
\hline
Trace input scope & 2 & 4 \\
\hline
Trace output scope & 1 & 4\\
\hline
Memo\_Table\_T entries & \multicolumn{2}{c|}{512}\\
\hline
Memo\_Table\_T assoc & \multicolumn{2}{c|}{4-way}\\
\hline
Memo\_Table\_G entries & n/a & 1024 \\
\hline
Memo\_Table\_G assoc & n/a & 4\\
\hline

\end{tabular}
\end{small}
\end{center}
\end{table}

\section{Reuse Inside Loops}
\label{sec:loops}

As loops execute many times the same instructions over a set of variables, they probably execute redundant instructions as input values are limited to a finite set. To test this hypothesis,  a subset of SPEC~2000~int benchmarks was analysed to verify the patterns of instruction reuse and how they are related to loops. For each benchmark, 150~million instructions were executed in {\tt sim-rst}. The last 100~thousand instructions were logged, with information about whether they were reused or not.

PC addresses of instructions that started traces were stored and then the log was followed backwards, finding common sequences of instructions occurring before these traces and how many times they occurred. These sequences characterize instructions inside loops. To be sure that these instructions were inside loops, the logs were verified for branch instructions with negative offsets, i.e., that jumped back to previous instructions. Preliminary results showed that a large number of traces were inside loops, hence we decided to investigate the contribution to performance of enabling reuse only inside traces.

Modifications in {\tt sim-rst} made it possible to enable reuse only inside or outside loops, which were dynamically identified by branches with backward addresses.

\subsection{Speedup}
%DONE
One of the main objectives of reusing instructions and traces is to improve performance. Hence, it is important to measure how much reusing traces created only inside loops will reduce performance improvements when compared to the non-speculative reuse mechanism DTM and the speculative reuse RST. A constrained version of RST with a single reuse table can achieve speedups of 1.24 over a baseline architecture without reuse~\cite{PIL07ijhpca}, and even more with separated tables for single instruction and multiple instruction traces. In this subsection, we compare RST-Loop's performance with those two alternative reuse schemes.

\begin{figure}[!ht]
 \centering\includegraphics[width=.5\textwidth]{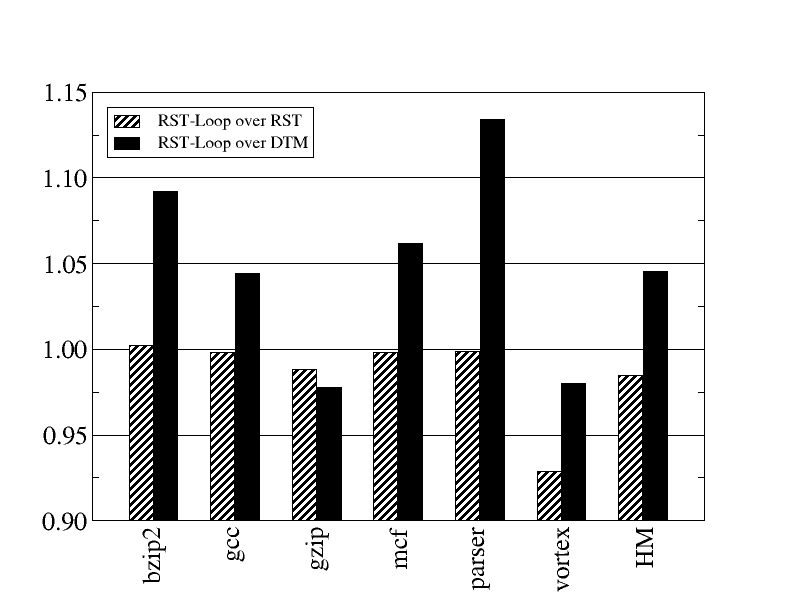}
\caption{Speedups of RST-Loop over RST and DTM}
\label{fig:speedup:loop}
%.985 and
\end{figure}

Figure~\ref{fig:speedup:loop} shows the speedup from RST-Loop over the original RST and DTM. The last column shows the harmonic mean~(HM) of the speedups. As RST-Loop restricts reuse to loops, it is to be expected that it presents less performance than the original RST. For most benchmarks, the performance hit was not statistically relevant. The only benchmark that presented a significant loss was {\tt vortex}, with 7.1\% less performance than with RST. The harmonic mean speedup for RST-Loop over RST showed only 1.5\% loss of performance.

When compared to DTM, which does not have speculative reuse, RST-Loop presented positive speedups for all benchmarks but {\tt gzip} and {\tt vortex}. Even for these two benchmarks the performance hit was small and less than 3\%. When compared to DTM, it becomes even more clear that RST~Loop can achieve most of the performance even if allowing reuse only inside loops, with a harmonic mean speedup of~1.045.

% \begin{table}[!ht]
%  \caption{Performance comparison between RST~Loop, RST and DTM}
% \label{tab:performance:loop}
% \begin{scriptsize}
% \begin{tabular}{l|c|c|c|c|c|c}
% & {\bf bzip2} & {\bf gcc} & {\bf gzip} & {\bf mcf} & {\bf parser} & {\bf vortex}\\
% \hline
% {\bf RST Loop/RST} &1.002 & 0.998 & 0.988 & 0.998 & 0.999 & 0.929\\
% \hline
% {\bf RST Loop/DTM} &1.092 & 1.044 & 0.978 & 1.062 & 1.134 & 0.980\\
% \hline
% \end{tabular}
% \end{scriptsize}
% \end{table}

\subsection{Number of Accesses to Reuse Table}

Each access to the reuse table consumes energy, therefore reducing the total number of accesses is expected to improve the overall energy consumption of processors with reuse mechanisms. In this subsection, we explore how RST-Loop and its complement, RST-Out-Of-Loop, behave in terms of accesses to the reuse tables when compared to the original RST.

\begin{figure}[!ht]
 \centering\includegraphics[width=.5\textwidth]{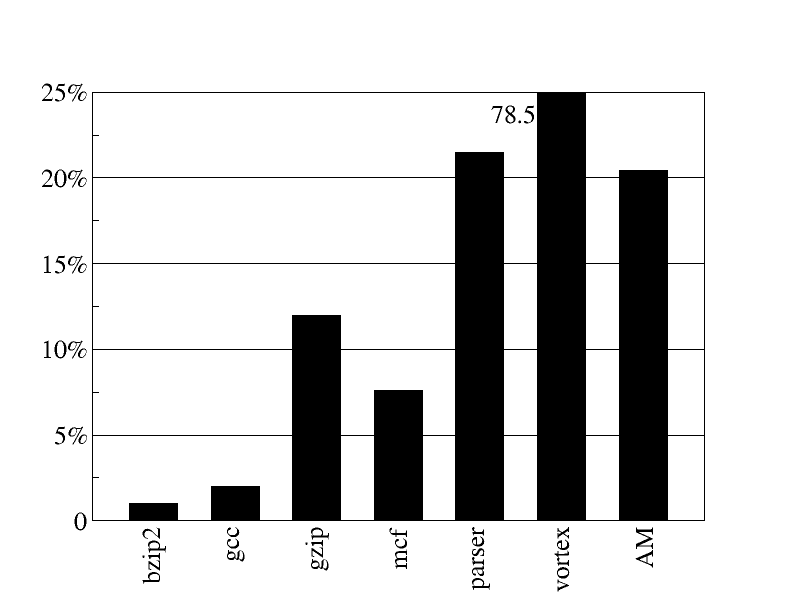}
\caption{Reduction on captured traces for RST-Loop and RST-Out-of-Loop (vortex).}
\label{fig:memoaccess:loop}
\end{figure}

Figure~\ref{fig:memoaccess:loop} shows the percentage of reduction in the number of captured traces when using RST-Loop and RST-Out-Of-Loop (for the {\tt vortex} benchmark only) compared to RST with unified reuse table~\cite{PIL07ijhpca}. Note that the bar representing {\tt vortex} is out of scale. In average, there was a reduction of 20.4\% in the number of accesses to the reuse table. All benchmarks presented less captured traces, but only {\tt mcf}, {\tt parser}, and {\tt vortex} showed significant reductions.

The benchmark {\tt vortex} is the one with the most interesting behaviour, with most of the reusable traces outside loops, which helps to explain its performance hit when reusing only inside loops (Figure~\ref{fig:speedup:loop}. RST-Loop was able to reduce the percent of table accesses in a range from 1 to 12\% for the benchmarks with good performance. In the specific case of {\tt vortex} and RST-Loop, it was able to achieve 98\% of the performance measured with non-speculative reuse~(DTM) while reducing {\em Memo\_Table\_T} access by 78.5\%, thus reducing energy consumption in a proportional way.

%
% \begin{table}[!ht]
% \caption{Comparison of captured traces in RST-Loop and RST-Out-Of-Loop against RST.}
% \label{tab:table:accesses}
% \begin{scriptsize}
% \begin{tabular}{c|c|c|c|c|c|c}
% \multicolumn{6}{c|}{{\bf Loop}} & {\bf Out-Of-Loop}\\
% \hline
%  bzip2 & gcc & gzip & mcf & parser & vortex & vortex \\
% \hline
% %Speedup &1.002 & 0.998 & 0.988 & 0.998 & 0.999 & 0.929 & 0.970\\
% %\hline
% -1.0\% & -2.0\% & -12.0\% & -7.6\%  & -1.6\% & -78.5\% & -21.5\%\\
% \hline
% \end{tabular}
% \end{scriptsize}
% \end{table}

\subsection{Memory Trade-offs}

To further study impact on performance and table access, three different in-chip memory trade-offs were simulated, all for the RST-Loop architecture with the exception of {\tt vortex}, where RST-Out-of-Loop was employed. The first configuration used the results from the previous experiments, simulating a 4-way, 512~entries reuse table, and 32~KB for each first level caches. Then, the budget for first level caches was doubled for the second configuration, while the reuse table size was kept the same. Finally, the last experiment uses the same amount of first level cache memory, but doubles the number of entries in the reuse table.

\begin{figure}[!ht]
\centering\includegraphics[width=.5\textwidth]{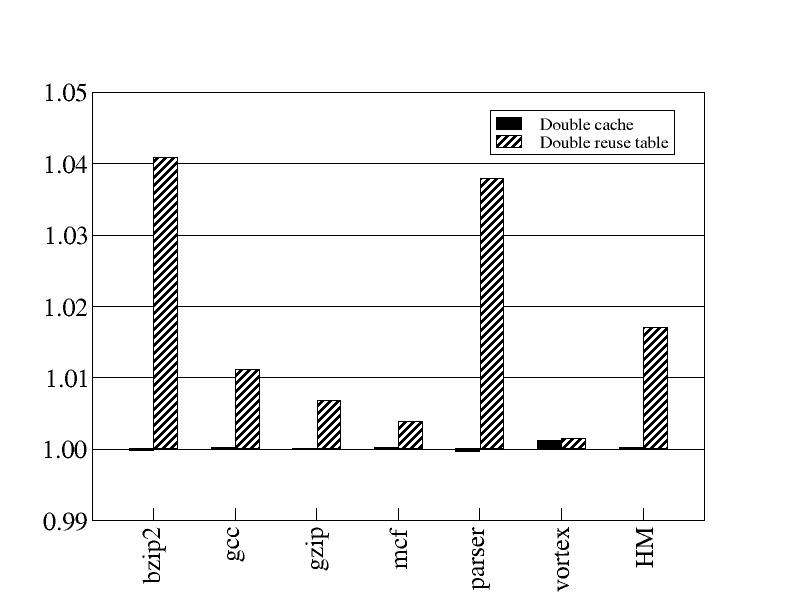}
 \caption{Speedups for memory trade-offs}
\label{fig:ipc:tradeoffs}
\end{figure}

\begin{figure}[!ht]
\centering\includegraphics[width=.5\textwidth]{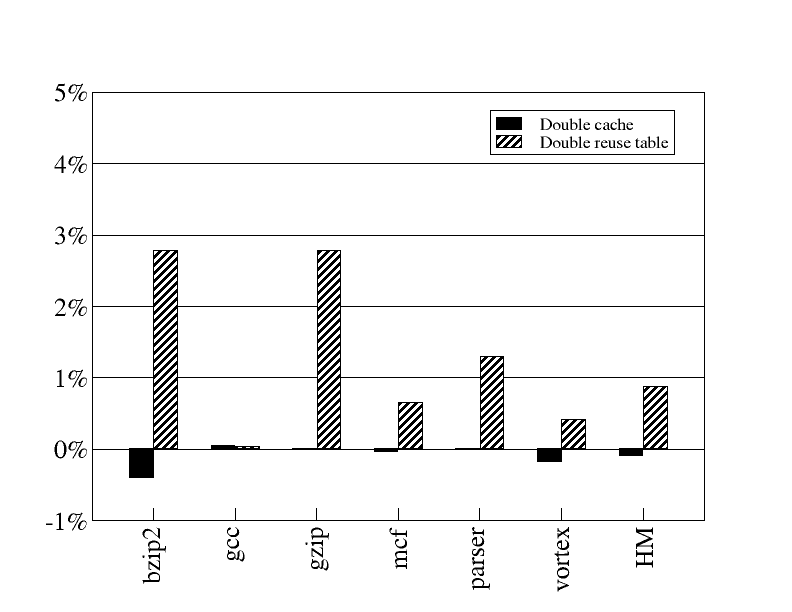}
 \caption{In-chip memory access difference for memory trade-offs.}
\label{fig:ipc:tradeoffs:access}
\end{figure}

Figure~\ref{fig:ipc:tradeoffs} shows the average difference of performance over the first configuration, while Figure~\ref{fig:ipc:tradeoffs:access} presents the difference between the number of in-chip memory accesses when compared to the first configuration. Doubling the first level cache budget does not improve performance, with an average speedup of only~1.02. The number of cache and reuse table accesses decreased only~0.09\%. Doubling the reuse table budget does not improve performance by much too, with an average speedup of 1.017. There was a decrease of 0.88\% in the total amount of cache and reuse table accesses, which cannot be considered significant. These results show that reuse may be an interesting alternative to larger cache budgets.

\section{Contribution from Different Instruction Types}
\label{sec:instructions}

In the next experiments, the contribution for reuse performance of different instruction types was studied. Former experiments with DTM and RST have restricted their reuse domain to instructions without side effects (like memory accesses and system calls) and that were not floating-point instructions. The former was due to the increased complexity of implementation. The later was done because floating-point instructions present less redundancy and require more storage space in the reuse tables.

Previous works considered a reuse domain that included all branches, integer logic-arithmetic instructions, and address calculations. For the measurement of the contribution of each subset of these instruction types, six different reuse domains were exploited: $B$ set, with only branches (conditional or not); $A$ set, with only add-subtract instructions (integer type); $M$ set, with only address calculations. The other subsets are the result of the difference between the $O$ original reuse domain set and the former three subsets: $\overline{B}$, $\overline{A}$, and $\overline{M}$.

Table~\ref{tab:distribution} shows the instruction type distribution by benchmark measured in our simulations of the baseline architecture. Notice that the column {\em Other} represents a very small percentage of the total.

\begin{table}[!ht]
\begin{center}
\caption{Distribution of instruction type by benchmark.}
\label{tab:distribution}
\begin{small}
 \begin{tabular}{|l|c|c|c|c|c|}
\hline
		& {\bf Memory} & {\bf Branches} & {\bf Logic and} & {\bf Other}\\
{\bf Benchmark} & {\bf Access} &		& {\bf Arithmetic} & 		\\
\hline\hline
{\bf bzip2} 	& 32.92\% 	& 14.46\%	& 51.33\%	   & 1.29\%\\
\hline
{\bf gcc}	& 39.32\%	& 20.09\%	& 39.98\%	   & 1.62\%\\
\hline
{\bf gzip}	& 30.44\%	& 15.32\%	& 50.97\%	   & 3.26\%\\
\hline
{\bf mcf}	& 34.59\%	& 28.16\%	& 36.74\%	   & 0.51\%\\
\hline
{\bf parser}	& 35.40\%	& 23.12\%	& 40.43\%	   & 1.05\%\\
\hline
{\bf vortex}	& 56.24\%	& 16.57\% 	& 26.84\%	   & 0.36\%\\
\hline
 \end{tabular}
\end{small}
\end{center}
\end{table}

Two main metrics were extracted from the results: the speedup, and the {\em Efficiency Index}~(EI). The Efficiency Index was developed specially for this work and compares the number of reused instructions by the speedup for a given subset of the reuse domain. It can be defined based on the speedup and the Reuse Rate $RR(S)$:
\[EI(S) = \frac{Speedup(S)}{RR(S)}\]
Where $S$ is one of the reuse domain subsets previously defined. The Efficiency Index correlates the potential of improving performance with each of the reuse subsets.

The Reuse Rate is defined as the fraction of instructions from the subset $S$ that have been reused:

 \[RR(S) = \frac{Reused\ instructions}{Executed\ instructions\ from\ S}\]

\subsection{Speedup}

\begin{figure}[!htp]
 \centering\includegraphics[width=.5\textwidth]{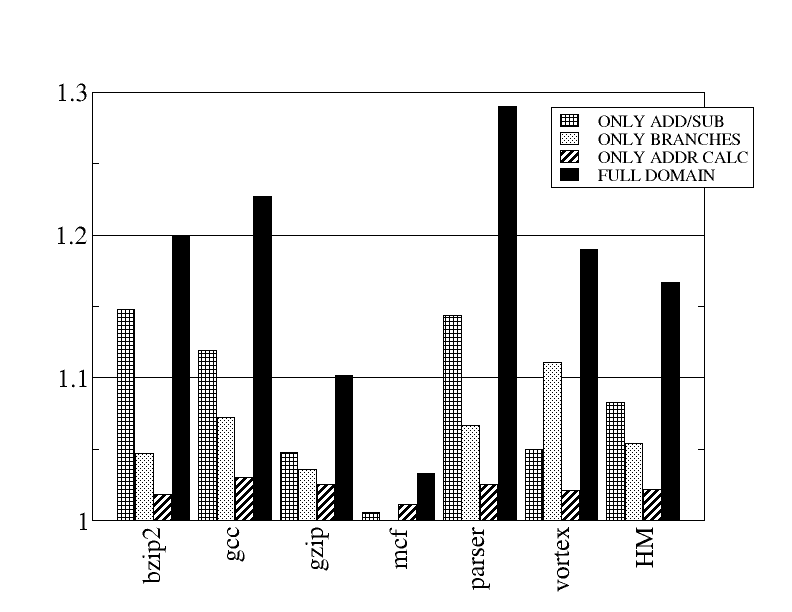}
 \caption{Harmonic mean speedups for selected subsets.}
\label{fig:speedup:all}
\end{figure}

\begin{figure}[!htp]
 \centering\includegraphics[width=.5\textwidth]{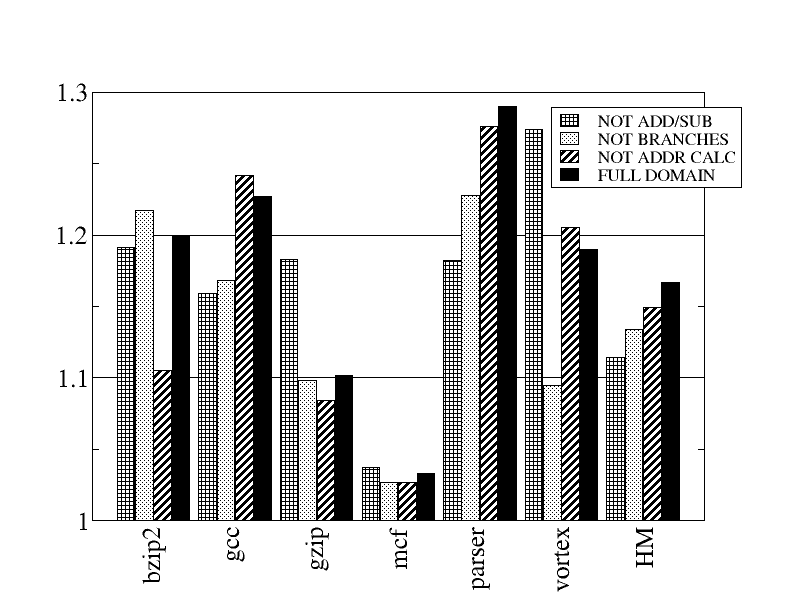}
 \caption{Harmonic mean speedups for complement of subsets.}
\label{fig:speedup:notall}
\end{figure}

As for the experiments with reuse inside and outside loops, performance is still an important metric when considering the different reuse domains.  Figure~\ref{fig:speedup:all} presents the harmonic mean speedups for the $B$, $A$, $M$, and $O$ domain reuse sets, while Figure~\ref{fig:speedup:notall} shows the same information for the complement subsets $\overline{A}$, $\overline{A}$, and $\overline{M}$. The last column of each graph presents the harmonic mean~(HM) for each set of instructions.

In Figure~\ref{fig:speedup:all}, the best performance is achieved with reuse of the full domain~$O$, with a speedup of~1.16 over the baseline. The subset $A$~(adds and subs) comes in second, with a speedup of~1.08. The subset $B$~(branches) comes in third, with a speedup of~1.05, while the subset $M$~(memory address calculation) comes in last place, with a speedup of~1.02. Most benchmarks showed the same behaviour, with only {\tt mcf} and {\tt vortex} presenting different sensibility to the reuse domain. Excluding full reuse domain, the benchmark {\tt mcf} was more sensible to reusing address calculations, while {\tt vortex} was more sensitive to branch reuse.

For the complementary sets shown in Figure~\ref{fig:speedup:notall}, where all instructions from the original reuse domain are reusable but for the selected subsets, the best results are obtained with full reuse domain, followed by not reusing address calculations~($\overline{M}$), then branches~($\overline{B}$), and finally add/sub~($\overline{A}$). The subset $\overline{B}$ presented a speedup of~1.13 over the baseline.

From these results, it is possible to assert that reusing only a limited subset of the instructions may still improve performance, but the benchmarks are very sensitive about which instructions are in the reuse domain. Restricting reuse may also reduce trace length, as instructions outside the reuse domain cannot be included in the trace and thus finish its building.

\subsection{Efficiency}

In this subsection, we discuss the efficiency of reusing each of the reuse domains as a function of the Efficiency Index as defined in the beginning of this Section. The EI presents how much a reuse domain improves performance in comparison with the number of reused instructions. A reuse domain is more efficient when it reuses less instructions but achieves a better speedup. Therefore, larger IE are better.

\begin{figure}[!htp]
\center

\subfloat[Selected subsets\label{fig:efficiency:all}]{\includegraphics[width=.5\textwidth]{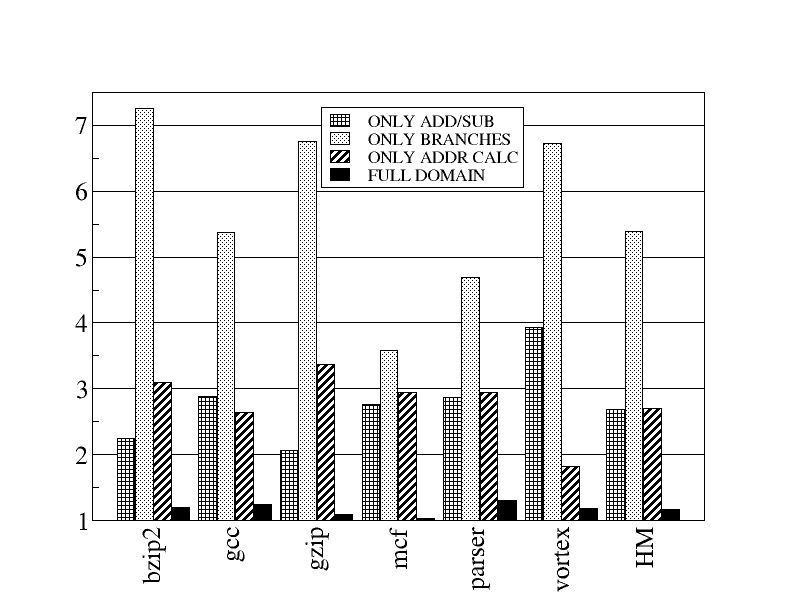}}
\hfill
\subfloat[Complement of subsets\label{fig:efficiency:notall}]{\includegraphics[width=.5\textwidth]{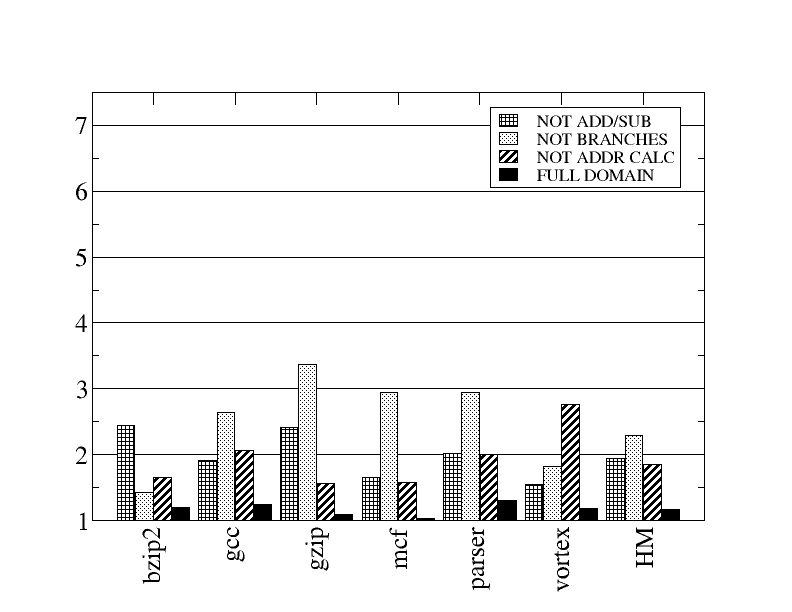}}
 \caption{Harmonic mean efficiency}
\label{fig:efficiency}
\end{figure}

Figure~\ref{fig:efficiency:all} shows the harmonic means of the Efficiency Index for the three subsets of the reuse domain, $B$, $A$, and $M$.  Figure~\ref{fig:efficiency:notall} shows the EI for the complement subsets. As expected, the original reuse domain (full domain) presents the lowest efficiency, although it presents the better performance improvements.

The best efficiency was achieved by branch instructions, where only 19.6\% of instructions were reused, but the speedup was 1.13. This is 90.3\% of the speedup found reusing all instructions. Therefore, we may assert that branch reuse is an important complement to branch prediction.

%FIXME Taxas de reuso?

An unexpected result is the contribution of add and subtract instructions to performance. Even though they are only 42.8\% of the reused instructions in average, reusing only these instructions allowed to achieve 92.8\% of the performance reusing the full domain. That may be explained by the small number of cycles that these instructions require for execution and how these impact less than branches and address calculations on the critical paths.

\subsection{Energy}
%TODO
Besides reducing the number of reuse table accesses, not reusing branches allows the reduction of the size of an entry in the trace reuse table {\em Memo\_Table\_T}~(Figure~\ref{fig:trace_memoization}). For the configurations simulated in this experiment, that would represent a decrease of 10.7\% in the area required for {\em Memo\_Table\_T} according to CACTI~\cite{THO:2008:cacti}. For configurations with smaller input and output scopes such as the ones simulated in the first set of experiments, this reduction would be even larger.

Simulation of the reuse table configurations using CACTI~5.2 and a~0.35~nm technology showed that the energy required to read a {\em Memo\_Table\_T} entry was 327.7~mW. The same entry without the fields for branch reuse would require~320.6~mW, a reduction of only~2\%. Therefore, reducing the number of accesses is more important than reducing the size of an entry in the table. As the subset including branches shows the best EI, leaving branch instructions outside of the reuse domain for such a small reduction in the energy spent in table reads would not compensate the performance loss.

\comentario{The newest CACTI can simulate at most 32nm. When we submitted the paper in January of 2013, 35nm was state-of-the art. We may simulate it again with 32nm if the reviewers insist, which is the most advanced technology CACTI 6.0 supports, although we do not think it will make much difference in the figures above. It is not advisable to simulate smaller processes with CACTI as it is not supported, and results would not be meaningful.}

\section{Conclusions and Future Work}
\label{sec:conclusion}

In this work, we presented a new evaluation of trace reuse taking into account the possibility of restricting reuse to some situations. Reusing only some classes of instructions or only instructions inside loops allow for reducing the amount of gate switching and required memory without losing much of the reuse potential. Hence, it becomes possible to apply trace reuse mechanisms in designs more restricted than general purpose superscalar processors, such as in the embedded market. Although \cite{Tsai:2011:vlsi} have already considered using reuse on embedded processors to lower power requirements, we understand that RST is a more flexible technique that can be tailored to different processors from embedded to general-purpose ones.

Studying the sensitivity of the trace reuse mechanism to instruction subsets makes it
possible to understand the contribution of each subset within the trace reuse context.
The creation of an Efficiency Index (EI), which is obtained from a division of each
subset speedup by the percentage of instructions executed in that subset, allows for a
more precise evaluation of the importance of each subset within the reuse mechanism.
Subsets of logical and arithmetic instructions, branch instructions and memory access
instructions were created and simulations were performed with them. We can conclude that
 logical and arithmetic instructions are very important in reuse mechanism, since
they alone produce 92.8\% of total reuse speedup, with only 42.8\% of the reusable rate.

An important result is that branch instructions have the best EI, since they alone produces 90.3\% of total reuse speedup, with only 14.5\% of reusable rate, even though a two-level branch predictor was used. Hence, further studies were reuse is focused on leverage branch prediction are highly desirable.  An unexpected result is the contribution of add and subtract instructions to performance. Even though they are only 42.8\% of the reused instructions, reusing only these instructions allowed to achieve 92.8\% of the performance reusing the full domain. Memory access instructions do not participate in trace reuse, only single instructions, but still produce more than twice the total reuse efficiency, using the index created as a comparative benchmark.

As more powerful processors are developed for smartphones and tables but still with strict power constraints, we intend to apply these results to the development of an architecture focused on high-end embedded processors.

\section*{Acknowledgments}

Work partially funded by grants from the Brazilian research agency CNPq.

\bibliographystyle{ws-jcsc}
\bibliography{ppd-abnt,navaux-abnt,common,arch}

\begin{thebibliography}{10}

\bibitem{ROT00micro}
A.~Roth and G.~S. Sohi, Register integration: A simple and efficient
  implementation of squash re-use, in {\em Proc. of the 33rd~Annual
  International Symposium on Microarchitecture\/},  (Los~Alamitos, IEEE
  Computer Society, Monterey, 2000), pp. 223--234.

\bibitem{SOD98micro}
A.~Sodani and G.~S. Sohi, Understanding the differences between value
  prediction and instruction reuse, in {\em Proc. of the 31st~Annual
  International Symposium on Microarchitecture\/},  (Los~Alamitos, IEEE
  Computer Society, 1998), pp. 205--215.

\bibitem{MOL99super}
C.~Molina, A.~Gonz{\'a}lez and J.~Tubella, Dynamic removal of redundant
  computations, in {\em Proc. of the 13th~ACM International Conference on
  Supercomputing\/},  (New~York, ACM, Rhodes, 1999), pp. 474--481.

\bibitem{HUA00comp}
J.~Huang and D.~J. Lilja, Extending value reuse to basic blocks with compiler
  support, {\em {IEEE} Transactions on Computers} {\bf 49}(April 2000)
  331--347.

\bibitem{COS99tr}
A.~T. da~Costa and F.~M.~G. Fran{\c{c}}a, The reuse potencial of trace
  memoization, Technical Report ES--498/99, {COPPE}--{UFRJ} (Rio de Janeiro,
  1999).

\bibitem{COS00pact}
A.~T. da~Costa, F.~M.~G. Fran{\c{c}a} and E.~M. {CHAVES FILHO}, The dynamic
  trace memoization reuse technique, in {\em Proc. of the 9th~International
  Conference on Parallel Architectures and Compilation Techniques\/},
  (Los~Alamitos, IEEE Computer Society, Philadelphia, October 2000), pp.
  92--99.

\bibitem{GON99icpp}
A.~Gonz{\'a}lez, J.~Tubella and C.~Molina, Trace-level reuse, in {\em Proc. of
  the 28th~International Conference on Parallel Processing\/},  (Los~Alamitos,
  IEEE Computer Society, Aizu-Wakamatsu, 1999), pp. 30--37.

\bibitem{HUA00pact}
J.~Huang and D.~J. Lilja, Exploring sub-block value reuse for superscalar
  processors, in {\em Proc. of the 9th~International Conference on Parallel
  Architectures and Compilation Techniques\/},  (Los~Alamitos, IEEE Computer
  Society, Philadelphia, October 2000), pp. 100--110.

\bibitem{WU01isca}
Y.~Wu, D.-Y. Chen and J.~Fang, Better exploration of region-level value
  locality with integrated computation reuse and value prediction, in {\em
  Proc. of the 28th~Annual International Symposium on Computer Architecture\/},
   (New~York, ACM, G{\"o}teborg, Sweden, June 2001), pp. 98--108.

\bibitem{OND01super}
S.~{\"O}nder and R.~Gupta, Load and store reuse using register file contents,
  in {\em Proc. of the 15th~ACM International Conference on Supercomputing\/},
  (New~York, ACM, Sorrento, Italy, 2001), pp. 289--302.

\bibitem{YAN00icpp}
J.~Yang and R.~Gupta, Load redundancy removal through instruction reuse, in
  {\em Proc. of the 29th~International Conference on Parallel Processing\/},
  (Los~Alamitos, IEEE Computer Society, Toronto, August 2000), pp. 61--68.

\bibitem{WAN97micro}
K.~Wang and M.~Franklin, Highly accurate data value prediction using hybrid
  predictors, in {\em Proc. of the 30th~Annual International Symposium on
  Microarchitecture\/},  (Los~Alamitos, IEEE Computer Society, December 1997),
  pp. 281--290.

\bibitem{SAT98micro2}
R.~Sathe, K.~Wang and M.~Franklin, Techniques for performing highly accurate
  data value prediction, {\em Microprocessors and Microsystems} {\bf
  22}(November 1998)  303--313.

\bibitem{SAZ97micro}
Y.~Sazeides and J.~E. Smith, The predictability of data values, in {\em Proc.
  of the 30th~Annual International Symposium on Microarchitecture\/},
  (Los~Alamitos, IEEE Computer Society, December 1997), pp. 248--258.

\bibitem{GRU98isca}
D.~Grunwald, A.~Klauser, S.~Manner and A.~Plezskun, Confidence estimation for
  speculation control, in {\em Proc. of the 25th~Annual International Symposium
  on Computer Architecture\/},  (New~York, ACM, Barcelona, June 1998), pp.
  122--131.

\bibitem{CAL99isca}
B.~Calder, G.~Reinman and D.~M. Tullsen, Selective value prediction, in {\em
  Proc. of the 26th~Annual International Symposium on Computer Architecture\/},
   (New~York, ACM, Atlanta, May 1999), pp. 64--74.

\bibitem{PIL03sbac}
M.~L. Pilla, P.~O.~A. Navaux, F.~M.~G. Fran{\c{c}}a, A.~T. da~Costa, B.~R.
  Childers and M.~L. Soffa, The limits of speculative trace reuse on deeply
  pipelined processors, in {\em Proc. of the 15th~Symposium on Computer
  Architecture and High-Performance Computing\/},  (S{\~a}o Paulo, SBC, S{\~a}o
  Paulo, October 2003), pp. 36--44.

\bibitem{HUA99tr}
J.~Huang, Y.~Choi and D.~Lilja, Improving value prediction by exploiting both
  operand and output value locality, Technical Report Technical Report ARCTic
  99-06, Laboratory for Advanced Research in Computing Technology and Compilers
  (July 1999).

\bibitem{PIL06sbac}
M.~L. Pilla, B.~R. Childer, A.~T. da~Costa, F.~M.~G. Fran{\c{c}}a and P.~O.~A.
  Navaux, A speculative trace reuse architecture with reduced hardware
  requirements, in {\em Proc.\ of the 18th~Symposium on Computer Architecture
  and High-Performance Computing\/},  eds. A.~F. de~Souza, R.~Buyya and W.~M.
  Jr. (Los Alamitos, IEEE Computer Society, Ouro Preto, October 2006), pp.
  47--54.

\bibitem{PIL07ijhpca}
M.~L. Pilla, B.~R. Childer, A.~T. da~Costa, F.~M.~G. Fran{\c{c}}a and P.~O.~A.
  Navaux, Limits for a feasible speculative trace reuse implementation, {\em
  International Journal of High Performance Systems Architecture} {\bf 1}(April
  2007)  69--76.

\bibitem{LIA02apccsa}
C.-H. Liao and J.-J. Shieh, Exploiting speculative value reuse using value
  prediction, in {\em Proc.\ of the 7th Asia-Pacific Conference on Computer
  Systems Architecture\/},  (Australian Computer Society, Inc., Melbourne,
  2002), pp. 101--108.

\bibitem{CONTRAIL2003}
T.~Koushiro, T.~Sato and I.~Arita, A trace-level value predictor for contrail
  processors, {\em SIGARCH Comput. Archit. News} {\bf 31}(3) (2003)  42--47.

\bibitem{Pratas:2008:ccf}
F.~Pratas, G.~Gaydadjiev, M.~Berekovic, L.~Sousa and S.~Kaxiras, Low power
  microarchitecture with instruction reuse, in {\em Proceedings of the 5th
  conference on Computing frontiers\/},  {\em CF '08}, (ACM, New York, NY, USA,
  2008), pp. 149--158.

\bibitem{Rotenberg:1996:micro}
E.~Rotenberg, S.~Bennett and J.~E. Smith, Trace cache: a low latency approach
  to high bandwidth instruction fetching, in {\em Proceedings of the 29th
  Annual {ACM/IEEE} {I}nternational {S}ymposium on {M}icroarchitecture\/},
  {\em MICRO 29}, (IEEE Computer Society, Washington, DC, USA, 1996), pp.
  24--35.

\bibitem{Tsumura:2007:pdcn}
T.~Tsumura, I.~Suzuki, Y.~Ikeuchi, H.~Matsuo, H.~Nakashima and Y.~Nakashima,
  Design and evaluation of an auto-memoization processor, in {\em Parallel and
  Distributed Computing and Networks\/},  ed. H.~Burkhart (IASTED/ACTA Press,
  2007), pp. 230--235.

\bibitem{Oda:2011:icnc}
R.~Oda, T.~Yamada, T.~Ikegaya, T.~Tsumura, H.~Matsuo and Y.~Nakashima, Input
  entry integration for an auto-memoization processor, in {\em ICNC\/},  (IEEE
  Computer Society, 2011), pp. 179--185.

\bibitem{Tsai:2011:vlsi}
Y.-Y. Tsai and C.-H. Chen, Energy-efficient trace reuse cache for embedded
  processors, {\em Very Large Scale Integration (VLSI) Systems, IEEE
  Transactions on} {\bf 19}(sept. 2011)  1681--1694.

\bibitem{Rahimi:2013:tcs}
A.~Rahimi, L.~Benini and R.~Gupta, Spatial memoization: Concurrent instruction
  reuse to correct timing errors in simd architectures, {\em Circuits and
  Systems II: Express Briefs, IEEE Transactions on} {\bf 60}(Dec 2013)
  847--851.

\bibitem{Rahimi:2014:date}
A.~Rahimi, L.~Benini and R.~K. Gupta, Temporal memoization for energy-efficient
  timing error recovery in gpgpus, in {\em Proceedings of the Conference on
  Design, Automation \& Test in Europe\/},  {\em DATE '14}, (European Design
  and Automation Association, 3001 Leuven, Belgium, Belgium, 2014), pp.
  100:1--100:6.

\bibitem{BUR97tr}
D.~C. Burger and T.~M. Austin, The {S}implescalar {T}ool {S}et, version 2.0,
  Technical Report CS-TR-1997-1342, University of Wisconsin (Madison, 1997).

\bibitem{THO:2008:cacti}
S.~Thoziyoor, N.~Muralimanohar, J.~H. Ahn and N.~P. Jouppi, {CACTI} 5.1, Tech.
  Rep. HPL-2008-20, {HP} Laboratories  (2008).

\bibitem{pact:00}
{\em Proc. of the 9th~International Conference on Parallel Architectures and
  Compilation Techniques} (Los~Alamitos, IEEE Computer Society, Philadelphia,
  October 2000).

\bibitem{micro:97}
{\em Proc. of the 30th~Annual International Symposium on Microarchitecture}
  (Los~Alamitos, IEEE Computer Society, December 1997).

\end{thebibliography}

\end{document}